\documentclass{jpconf}
\usepackage{graphicx}
\usepackage{amsmath}

\newcommand{\sss}[1]{{\scriptscriptstyle{#1}}}

\newcommand{\nS}{n_{_{\mathrm{S}}}}
\newcommand{\alphaS}{\alpha_{_{\mathrm{S}}}}

\newcommand{\usssS}{\mathrm{\sss{S}}}

\newcommand{\ue}{\mathrm{e}}

\newcommand{\us}{\mathrm{s}}

\newcommand{\cs}{c_\us}

\newcommand{\calP}{\mathcal{P}}

\newcommand{\mpl}{m_{_\mathrm{Pl}}}

\newcommand{\order}[1]{\mathcal{O}\!\left(#1\right)}

\newcommand{\deriv}[2]{#1_{\negthinspace,#2}}
\newcommand{\dderiv}[3]{#1_{\negthinspace,#2#3}}

\newcommand{\ud}{\mathrm{d}}

\newcommand{\qmode}{\mu_\usssS}
\newcommand{\piv}{\diamond}

\newcommand{\etaS}{\eta_*}

\newcommand{\etaP}{\eta_\piv}
\newcommand{\qmodeS}{\mu_\usssS}

\newcommand{\kpiv}{k_\piv}
\newcommand{\epsoneP}{\epsilon_{1\piv}}
\newcommand{\epstwoP}{\epsilon_{2\piv}}
\newcommand{\epsthreeP}{\epsilon_{3\piv}}

\newcommand{\deltaoneP}{\delta_{1\piv}}
\newcommand{\deltatwoP}{\delta_{2\piv}}

\newcommand{\csP}{\cs {}_\piv}
\newcommand{\epsoneS}{\epsilon_{1*}}
\newcommand{\epstwoS}{\epsilon_{2*}}

\newcommand{\deltaoneS}{\delta_{1*}}

\begin{document}
\title{Dirac-Born-Infeld and k-inflation: the CMB anisotropies from
  string theory}

\author{Christophe Ringeval}

\address{Theoretical and Mathematical Physics Group, Centre for
  Particle Physics and Phenomenology, Louvain University, 2 Chemin du
  Cyclotron, 1348 Louvain-la-Neuve, Belgium}

\ead{christophe.ringeval@uclouvain.be}

\begin{abstract}
  Inflationary models within string theory exhibit unusual scalar
  field dynamics involving non-minimal kinetic terms and generically
  referred to as k-inflation. In this situation, the standard
  slow-roll approach used to determine the behavior of the primordial
  cosmological perturbations cannot longer be used. We present a
  generic method, based on the uniform approximation, to analytically
  derive the primordial power spectra of scalar and tensor
  perturbations. At leading order, the scalar spectral index, its
  running and the tensor-to-scalar ratio are modified by the new
  dynamics. We provide their new expression, correct previous results
  at next-to-leading order and clarify the definition of what is the
  tensor-to-scalar ratio when the sound horizon and Hubble radius are
  not the same. Finally, we discuss the constraints the parameters
  encoding the non-minimal kinetic terms have to satisfy, such as the
  sound speed and the energy scale of k-inflation, in view of the
  fifth year Wilkinson Microwave Anisotropy Probe (WMAP5) data.
\end{abstract}

\section{Introduction}

In the context of string theory, cosmic inflation can be achieved
through the motion $D$-branes in higher dimensional warped and compact
spacetimes~\cite{Kachru:2003sx}. There, the inflaton appears as the
scalar degree of freedom associated with the position of a brane in
these extra-dimensions. From a four-dimensional point of view, Lorentz
invariance along a $D$-brane necessarily leads to four-dimensional
scalar fields exhibiting non-standard kinetic terms, and more
precisely of the Dirac--Born--Infeld (DBI) type~\cite{Dvali:1998pa,
  Kachru:2003aw, Langlois:2008wt}. In fact, DBI-inflation belongs to
the class of k-inflationary models in which the accelerated expansion
of the universe can be driven by the scalar field's kinetic terms
instead of its potential
energy~\cite{ArmendarizPicon:1999rj}. Assuming the gravity sector to
be described by General Relativity, the action of the ''k-inflaton''
$\varphi(x^\mu)$ reads
\begin{equation}
  S=\dfrac{1}{2\kappa } \int \ud^4x \sqrt{-g} \left[ R + 2\kappa
    P\left(X,\varphi\right) \right],
\end{equation}
where $\kappa \equiv 8\pi/\mpl^2$ and $X \equiv - g^{\mu \nu}
\partial_\mu \varphi \partial_\nu \varphi /2$. The quantity
$P(X,\varphi)$ is any \emph{acceptable} functional of $X$ and
$\varphi$ (see Ref.~\cite{Bruneton:2007si}).  All kinetically modified
inflationary models have a ``speed of sound''
\begin{equation}
\label{eq:csdef}
\cs^2 \equiv \dfrac{\deriv{P}{X}}{\deriv{P}{X} + 2 X \dderiv{P}{X}{X}}\,,
\end{equation}
which is generically different from the speed of light. In a flat
Friedmann-Lema\^{\i}tre--Robertson--Walker (FLRW) universe, it was
shown in Ref.~\cite{Garriga:1999vw}, that the Mukhanov--Sasaki mode
function $\qmode (k,\eta)= \zeta \sqrt{2\kappa \epsilon_1}/\cs$
($\zeta$ being the comoving curvature perturbation) satisfies the
modified equation of motion
\begin{equation}
\label{eq:qmode}
\qmodeS'' + \left[\cs^2(\eta) k^2 - \frac{\nu^2(\eta)-1/4}{\eta
    ^2}\right]
\qmodeS = 0.
\end{equation}
The effective potential is $(\nu^2-1/4)/\eta^2 = [\ln (a
\sqrt{\epsilon_1}/\cs)]''$, and all derivatives are with respect to
conformal time $\eta$. The quantity $a(\eta)$ stands for the FLRW
scale factor while $\epsilon_1 = -\ud \ln H/\ud \ln a$ is the first
Hubble flow function ($H$ being the Hubble parameter). The standard
form of this equation is recovered by setting $\cs=1$ and can be
solved by defining a hierarchy of Hubble flow functions encoding the
rate of change of the Hubble parameter and its higher order
logarithmic derivatives: $\epsilon_{n+1} \equiv \ud \ln
|\epsilon_n|/\ud \ln a$. Assuming slow-roll, i.e. $\epsilon_n \ll 1$,
one can expand the effective potential $(\nu^2-1/4)/\eta^2$ and solve
order by order Eq.~(\ref{eq:qmode}) along the lines of
Refs.~\cite{Schwarz:2001vv, Leach:2002ar,
  Schwarz:2004tz}. Generalising this method to the k-inflationary case
in which $\cs(\eta)$ is not constant requires some care. Indeed, both
the effective potential and the propagation speed are modified. In the
following, we use the uniform approximation to solve
Eq.~(\ref{eq:qmode}) order by order to predict the shape of the tensor
and scalar primordial fluctuations at the origin of the CMB
anisotropies.

\section{K-inflationary perturbations}


For k-inflation, we can define a new hierarchy encoding the rate of
change of the sound speed, the sound flow functions $\delta_{i}$,
defined by
\begin{equation}
\label{eq:defdels}
\delta_{n+1}=\frac{\ud \ln \vert \delta_n \vert}{\ud \ln a}\,, \qquad
\delta_{0} \equiv \dfrac{{\cs}_{\mathrm{in}}}{\cs}\,.
\end{equation}
Expanding both the sound speed and the effective potential in terms of
the Hubble and sound flow functions around a particular conformal time
$\etaS$ gives~\cite{Lorenz:2008et}
\begin{equation}
\nu^2(\eta)  =\dfrac{9}{4} + 3 \epsoneS + \dfrac{3}{2} \epstwoS + 3
\deltaoneS + \order{\epsilon \delta} = \nu_*^2 + \order{\epsilon
  \delta},\quad  
\cs(\eta)  =\cs {}_*
\left(1 + \deltaoneS \ln\dfrac{\eta}{\etaS}
\right) + \order{\epsilon \delta} ,
\end{equation}
where all stars mean that the corresponding function is evaluated at
$\etaS$. From these expressions, one can solve Eq.~(\ref{eq:qmode}) at
first order in the flow functions $\epsilon_i$ and $\delta_i$ by using
the uniform approximation~\cite{Habib:2002yi, Habib:2004kc}.

The scalar primordial power spectrum, at first order in Hubble and
sound flow functions, then reads~\cite{Lorenz:2008et}
\begin{equation}
\label{eq:scalarspectrum}
{\cal P}_{\zeta} = \dfrac{H_\piv^2}{\pi
    \mpl^2\epsoneP\csP} \Biggl[1 - 2(D+1) \epsoneP - D \epstwoP +
  (D+2) \deltaoneP - \left(2\epsoneP + \epstwoP -\deltaoneP \right)
  \ln \dfrac{k}{k_\piv} \Biggr] ,
\end{equation}
where $D=1/3-\ln 3$. All diamond indexed quantities are evaluated at
the time $\etaP$ defined to be the time at which a chosen pivot
wavenumber $\kpiv$ crossed the sound horizon during inflation,
i.e. the solution of $-\kpiv \etaP = 1/\csP$. The constant factor $18
\ue^{-3}$ typical of WKB methods has also been absorbed in the
definition of $H_\piv$~\cite{Martin:2002vn}.

Concerning the tensor modes, their evolution is not affected by the
non-standard kinetic terms and their power spectrum remains the same
as in the standard case $\cs=1$. However, an important, and so far
overlooked, difference is that the standard tensor power spectrum is
evaluated at the time at which the perturbations crossed the Hubble
radius during inflation. As a result, it is expressed in terms of the
Hubble flow functions evaluated at a different time than $\etaP$ and
one cannot evaluate a tensor-to-scalar ratio by simply dividing both
power spectra. Using the Hubble and sound flow expansion, one can
nevertheless express the tensor power spectrum at $\etaP$. After some
algebra, one obtains
\begin{equation}
\label{eq:tensorspectrum}
\calP_h(k)  = \dfrac{16 H_\piv^2}{\pi \mpl^2}  \Biggl[1 - 2\left(D + 1 
  - \ln \csP\right) \epsoneP - 2 \epsoneP \ln \dfrac{k}{k_\piv} \Biggr].
\end{equation}
We immediately see that the speed of sound influences $\calP_h$, and
this effect becomes all the more so important than $\cs$ is small. The
above expression explains the numerical results on the
tensor-to-scalar ratio discussed in Ref.~\cite{Agarwal:2008ah}.


From Eqs.~(\ref{eq:scalarspectrum}) and (\ref{eq:tensorspectrum}), one
can deduce the scalar spectral index $\nS$, its running $\alphaS$ and
the tensor-to-scalar ratio at next-to-leading
order~\cite{Schwarz:2004tz,Lorenz:2008et}
\begin{equation}
\begin{aligned}
  \nS-1 \equiv \left(\frac{\ud\ln\mathcal{P}_\zeta}{\ud\ln k}\right)
  _{k=k_\piv} & = -2\epsoneP-\epstwoP+\deltaoneP-2\epsoneP^{2} -
  (2D+3) \epsoneP\epstwoP +3\epsoneP\deltaoneP + \epstwoP\deltaoneP \\
  & - D \epstwoP\epsthreeP-\deltaoneP^{2}  + (D+2)  \deltaoneP\deltatwoP, \\
  \alphaS \equiv \left(\frac{\ud^{2}\ln\mathcal{P}_\zeta}{\ud\ln^{2}
      k}\right)_{k=k_\piv} & = -2\epsoneP \epstwoP - \epstwoP
  \epsthreeP
  +  \deltaoneP \deltatwoP, \\
  r \equiv \left.\dfrac{\calP_h}{\calP_\zeta}\right|_{k=\kpiv} & = 16
  \csP \epsoneP \left[ 1+ 2 \epsoneP \ln \csP + D \epstwoP - (D+2)
    \deltaoneP \right].
\end{aligned}
\end{equation}
The spectral index and running correct previous results at
next-to-leading order~\cite{Peiris:2007gz, Bean:2007hc,
  HenryTye:2006uv}, which were assuming $\cs$ constant, and match with
another method proposed by Kinney and Tzirakis in
Ref.~\cite{Kinney:2007ag}. The term in $\ln\csP$ in the
tensor-to-scalar ratio has to be considered as soon as $\csP$ becomes
small enough.

\section{Conclusion}

From the scalar and tensor power spectra given above, one can compare
the predicted CMB anisotropies with the current data. In
Ref.~\cite{Lorenz:2008je}, we have performed a
Monte--Carlo--Markov--Chains analysis of the WMAP5
data~\cite{Komatsu:2008hk} against the k-inflationary power
spectra. At $95\%$ of confidence, the flow parameters and the energy
scale of k-inflation have to verify
\begin{equation}
\label{eq:nstwosig}
0.003 \le 2 (\epsoneP-\deltaoneP) + (\epstwoP + \deltaoneP) \le 0.075,
\quad \log (\epsoneP \csP) \le -2.3, \quad \ln \left(10^5 \dfrac{H_\piv}{\mpl} \right) \le -0.59 .
\end{equation}
Notably, due to the new degree of freedom introduced by $\cs$, we do
not longer find any bound on $\epsoneP$ alone. The class of
k-inflationary models is thus weakly constrained by CMB data. Let us
however mention that the subclass of DBI models generate a large
amount of non-Gaussianity when $\cs$ becomes
small~\cite{Chen:2006nt}. In this later case, one can show that the
current WMAP5 bounds on non-Gaussianity imposes that $\log \epsoneP
\le -1.1$, at two-sigma~\cite{Lorenz:2008je}.

\ack It's a pleasure to thank L.~Lorenz and J.~Martin for their
comments on the manuscript. This work is supported by the Belgian
Federal Office for Science, Technical and Cultural Affairs, under the
Inter-university Attraction Pole grant P6/11.

\section*{References}
\bibliography{references}
\end{document}